\documentclass[technote]{IEEEtran}

\usepackage{subcaption}
\usepackage{graphicx}
\usepackage{todonotes}
\usepackage{titlesec}         
\DeclareMathSizes{10}{9}{6}{5}
\usepackage{lipsum} 
\usepackage{cite}
\usepackage{xspace}
\usepackage{subcaption}
\usepackage[cmex10]{amsmath}
\usepackage{mathtools,amssymb,lipsum}
\usepackage{lettrine}
\usepackage{bm}
\usepackage[linesnumbered,ruled,lined]{algorithm2e}
\usepackage{array}
\usepackage{url}  
\usepackage{amsthm}
\usepackage{multirow}
\usepackage{color}
\usepackage{epstopdf}
\usepackage{endnotes}
\usepackage{framed}
\usepackage{cool}
\usepackage{enumerate}
\usepackage{url}
\usepackage[subnum]{cases}
\usepackage[nocomma]{optidef}
\usepackage{etoolbox}
\usepackage{amsmath,amssymb,amsfonts}
\usepackage{textcomp}

\begin{document}
\title{\Large{Joint Beamforming and Trajectory Optimization for Multi-UAV-Assisted Integrated Sensing and Communication Systems}}
\author{Yan~Kyaw~Tun,~\IEEEmembership{Member,~IEEE}, Nway~Nway~Ei, Sheikh~Salman~Hassan,~Cedomir~Stefanovic,~\IEEEmembership{Senior Member,~IEEE}, 
Nguyen~Van~Huynh,~\IEEEmembership{Member,~IEEE},
Madyan~Alsenwi, and Choong~Seon~Hong,~\IEEEmembership{Fellow,~IEEE}
\thanks{Yan Kyaw Tun and Cedomir Stefanovic are with the Department of Electronic Systems, Aalborg University, 2450 København SV, Denmark, Email:{\{ykt, cs\}@es.aau.dk}.}
\thanks{Sheikh Salman Hassan is with the Institute for Imaging, Data, and Communications, The University of Edinburgh, EH9 3FG Edinburgh, UK, Email: shassan@ed.ac.uk}
\thanks{Nway Nway Ei and Choong Seon Hong are with the Department of Computer Science and Engineering, Kyung Hee University, Yongin, 17104, Republic of Korea, Emails:{\{nwayei, cshong\}@khu.ac.kr}. }
\thanks{N. V. Huynh is with the Department of Electrical Engineering
and Electronics, University of Liverpool, Liverpool, L69 3GJ, UK, E-mail: huynh.nguyen@liverpool.ac.uk. }
\thanks{Madyan Alsenwi is with the University of Luxembourg, E-mail: madyan.alsenwi@uni.lu.}
} 
\maketitle 
\begin{abstract}
In this paper, we investigate beamforming design and trajectory optimization for a multi-unmanned aerial vehicle (UAV)-assisted integrated sensing and communication (ISAC) system. The proposed system employs multiple UAVs equipped with dual-functional radar-communication capabilities to simultaneously perform target sensing and provide communication services to users. We formulate a joint optimization problem that aims to maximize the sum rate of users while maintaining target sensing performance through coordinated beamforming and UAV trajectory design. To address this challenging non-convex problem, we develop a block coordinated descent (BCD)-based iterative algorithm that decomposes the original problem into tractable subproblems. Then, the beamforming design problem is addressed using fractional programming, while the UAV trajectory is refined through the deep deterministic policy gradient (DDPG) algorithm. The simulation results demonstrate that the proposed joint optimization approach achieves significant performance improvements in both communication throughput and sensing accuracy compared to conventional, separated designs. We also show that proper coordination of multiple UAVs through optimized trajectories and beamforming patterns can effectively balance the tradeoff between sensing and communication objectives.       
\end{abstract}
\section{Introduction and Motivation}
The rapid advancement of wireless networks has highlighted the limitations of terrestrial infrastructure, particularly in remote areas, necessitating innovative solutions. Unmanned aerial vehicles (UAVs), with their mobility and flexible deployment capabilities, have emerged as promising enablers for aerial-terrestrial integrated networks (ATINs) due to their low latency and favourable propagation characteristics \cite{UAVs_networks}. Traditionally, UAVs equipped with radar sensors perform communication and sensing functions separately, using dedicated frequency bands. However, this approach leads to inefficient spectrum utilization and increased hardware complexity. To address these limitations, integrated sensing and communication (ISAC) has emerged as a compelling paradigm that enables simultaneous radar sensing and communication using shared spectrum and hardware resources.

The application of ISAC in UAV platforms presents unique opportunities and challenges with respect to the terrestrial systems.
The 
opportunities include improved spectrum efficiency, reduced hardware size, and lighter payloads.
However, UAVs face challanges such as limited onboard computing resources, which can make complex radar data analysis computationally burdensome.
Despite these constraints, ISAC can enable 
critical functionalities for UAV operations, like collision avoidance through sensing and facilitating information exchange via the Internet of aerial vehicles.

Recent advances in multi-antenna technologies and beamforming techniques have enabled more sophisticated ISAC implementations \cite{10410213}. However, the joint design of transmit beamforming and UAV trajectory optimization in multi-UAV scenarios presents several unique challenges \cite{10327740}. First, the spatial degrees of freedom must be carefully managed to balance the conflicting objectives of radar sensing and communication. Second, the mobility of multiple UAVs introduces additional complexity in coordinating their trajectories while maintaining effective coverage \cite{9804341}. Third, the coupling between beamforming designs and UAV positions creates a non-convex optimization problem that requires efficient solution methods.
This paper addresses these challenges by proposing a comprehensive framework for joint transmit beamforming and trajectory optimization in multi-UAV ISAC systems. Our key contributions are summarized as follows:
\begin{itemize}
    \item We design a system model that captures the interaction between multiple UAVs performing dual-functional radar-communication operations.
    \item We formulate a joint optimization problem to maximize the sum rate of the communication users while adhering to quality-of-service and sensing performance constraints.
    \item We propose an alternating optimization approach that uses fractional programming and deep deterministic policy gradient (DDPG) algorithm to solve the beamforming and UAV trajectory optimization subproblems, respectively.
    \item Our simulation results validate the effectiveness of the proposed method in improving sensing accuracy and communication rates compared to existing benchmarks.
\end{itemize}

\section{System Model}
This paper considers a multi-UAV-assisted ISAC system. In the proposed system model, a set $\mathcal{U}$ of $U$ UAVs equipped with $M$-uniform linear antenna arrays fly from their initial to final locations over a period $T$. During the flight period, each UAV $u$ communicates with a set $\mathcal{V}_u$ of $V_u$ ground users (GUs) and senses a set $\mathcal{K}_u$ of $K_u$ targets on demand, where $\cup_{u=1}^{U} V_u = V, V_u \cap V_{u'} = \emptyset, u \neq u'$, 
and $\cup_{u=1}^{U} K_u = K, k_u \cap k_{u'} = \emptyset, u \neq u', \forall u, u' \in \mathcal{U}$. For simplicity, we divide the 
flight period $T$ into discrete time slots $N$, where each time slot $n$ has the duration of $\tau = \frac{T}{N}$. Thus, we define $o_u[n] =(x_u[n], y_u[n], H_u[n])$ as the 3D coordinates of UAV $u$ at time slot $n$. Furthermore, let $l_v[n] =(x_v[n], y_v[n], 0)$, and  $l_k[n] =(x_k[n], y_k[n], 0)$ be the 3D coordinates of user $v$, and target $k$ at time slot $n$. 
\subsection{Communication Model}

We assume that distinct waveforms are employed for communication and sensing purposes, due to their different performance requirements (i.e., high data rates vs high correlation, respectively), 
and adopt the probabilistic line-of-sight (LoS) channel model for both communication and target sensing \cite{tun2022collaboration}.
Therefore, the probability of existing LoS link between UAV $u$ and user $v$ or target $k$ at time slot $n$ can be expressed as: 
\begin{equation}
    {\Pr}^{u,\textrm{LoS}}_{{v (\textrm{or})k}, n}= \frac{1}{1 + C \exp\bigg[-D(\phi_{v(\textrm{or})k,n}^u-C)\bigg]},
\end{equation}
where $\phi_{v(\textrm{or})k,n}^u = \frac{180}{\pi}\sin^{-1}(\frac{H_{u}[n]}{d_{v (\textrm{or})k,n}^u })$ with $d_{v (\textrm{or})k,n}^u = \sqrt{ \left\lVert (o_u[n] - l_{v (\textrm{or})k} [n]\right\rVert^2 +(H_u[n])^2}$, $C$ and $D$ are the constants that depend on the operating environment. Then, the probability of an existing non-LoS (NLoS) link between UAV $u$ and user $v$ or target $k$ at time slot $n$ can be expressed as $ {\Pr}^{u,\textrm{NLoS}}_{{v (\textrm{or})k}, n} = 1 -{\Pr}^{u,\textrm{LoS}}_{{v (\textrm{or})k}, n}$. As a result, the achievable channel gain between UAV $u$ and user $v$ at time slot $n$ can be formulated as:
\begin{equation}
\begin{split}
    \boldsymbol{h}_v^u[n] &= \sqrt{{\Pr}^{u,\textrm{LoS}}_{v,n} \alpha_0 (d_{v,n}^u)^{-2}} \boldsymbol{h}_v^{u,\textrm{LoS}}[n]  \\
          &  + \sqrt{{\Pr}^{u,\textrm{NLoS}}_{v,n} \alpha_0 (d_{v,n}^u)^{-2}} \boldsymbol{h}_v^{u,\textrm{NLoS}}[n], \forall u, \forall v, \forall n, 
    \end{split}
\end{equation}
where $\alpha_0$ is the path loss at reference distance $d_0 = 1$ m, $\boldsymbol{h}_v^{u,\textrm{LoS}}[n]= \boldsymbol{s}_v^u[n] = [1, e^{-j \frac{2\pi d}{\lambda}\sin(\phi_{v,n}^u)}, \dots,  e^{-j \frac{2\pi (M-1)d}{\lambda}\sin(\phi_{v,n}^u)}]^T$ is the transmit steering vector, where $\lambda$ is the wavelength, and $d$ stands for the antenna separation. Furthermore, the complex Gaussian random vector $\boldsymbol{h}_v^{u,\textrm{NLoS}}[n]$ has zero mean and unit covariance matrix. Finally, the achievable data rate of user $v$ associated with UAV $u$ at time slot $n$ can be expressed as:
\begin{equation}
   R_v^u[n] = \tau \log_2 \left( 1 + \frac{| (\boldsymbol{h}_v^u[n])^\dagger \boldsymbol{g}_v^u [n]|^2}{I^{\textrm{intra}} + I^{\textrm{inter}} + \sigma^2 }  \right),  \forall u, \forall v, \forall n,
\end{equation}
where $(.)^\dagger$ represents the conjugate transpose operation, $\sigma^2$ is the Additive white Gaussian noise (AWGN) power, and $I^{\textrm{intra}} = |\sum_{v'=1, v' \neq v }^{V_u} (\boldsymbol{h}_{v}^u[n])^\dagger \boldsymbol{g}_{v'}^u[n] + \sum_{k=1}^{K_u} (\boldsymbol{h}_{v}^u[n])^\dagger \boldsymbol{i}_{k}^u[n]|^2$ and $I^{\textrm{inter}} = | \sum_{u'=1, u' \neq u}^{U}\sum_{v'=1}^{V_{u'}} (\boldsymbol{h}_{v}^{u'}[n])^\dagger \boldsymbol{g}_{v'}^{u'}[n] + \sum_{u'=1, u' \neq u}^{U} \sum_{k=1}^{K_{u'}} (\boldsymbol{h}_{v}^{u'}[n])^\dagger \boldsymbol{i}_{k}^{u'}[n]|^2$ are the intra- and inter-interference at user $v$ associated with UAV $u$, respectively.
\subsection{Sensing Model}
Each UAV transmits signals, which are reflected off the targets and returned to the UAV. The Doppler shifts resulting from the movement of both targets and the UAV are assumed to remain constant throughout a given time slot and can be effectively compensated \cite{liu2018mimo}. Thus, the channel gain of the echo signal of target $k$ at UAV $u$ can be formulated as \cite{yuan2021integrated}:
\begin{equation}
\begin{split}
    \boldsymbol{\varphi}_k^u[n] &= \beta_{k,n}^u\sqrt{{\Pr}^{u,\textrm{LoS}}_{k,n}}  \boldsymbol{\varphi}_k^{u,\textrm{LoS}} [n]  + \beta_{k,n}^u \sqrt{{\Pr}^{u,\textrm{NLoS}}_{k,n} \kappa}  \\
    &\ \ \ \ \ \ \ \ \ \ \ \ \ \ \ \ \ \ \ \ \ \ \ \ \ \ \ \ \boldsymbol{\varphi}_k^{u,\textrm{NLoS}} [n], \forall u, \forall k, \forall n, 
    \end{split}
\end{equation}
where $\kappa$ is the LoS parameter, $\beta_{k,n}^u = \frac{\sigma_k}{2d_{k,n}^u}$ in which $\sigma_k$ is the radar cross-section \cite{yuan2021integrated}, $\boldsymbol{\varphi}_k^{u,\textrm{LoS}} [n] = \boldsymbol{a}_{k,n}^u (\boldsymbol{a}_{k,n}^u)^H$ where $\boldsymbol{a}_{k,n}^u = [1, e^{-j \frac{2\pi d}{\lambda}\sin(\phi_{k,n}^u)}, \dots,  e^{-j \frac{2\pi (M-1)d}{\lambda}\sin(\phi_{k,n}^u)}]^T$ is the transmit steering vector, and $ \boldsymbol{\varphi}_k^{u,\textrm{NLoS}}[n] $ is the complex Gaussian random vector with zero mean and unit covariance matrix. After receiving the reflected or echo signals, each UAV estimates the parameters of the targets. The Cramér-Rao bound (CRB) is a crucial metric for parameter estimators in radar sensing, providing a lower bound on the parameter estimators' mean square error (MSE) \cite{liu2021cramer}. Thus, for UAV $u$ at time slot $n$ where the directions of other targets are known, the CRB for estimating the parameter $\phi_{k,n}^u$ of target $k$ is given by \cite{de2009code}:
\begin{equation}
    \text{CRB} (\phi_{k,n}^u)  = \frac{\sigma^2}{2 |\beta_{k,n}^u|^2 (\Tr((\overline{A_{k,n}^u})^\dagger \overline{A_{k,n}^u} \boldsymbol{i_k^u}[n] (\boldsymbol{i_k^u}[n])^\dagger))},
\end{equation}
where $\overline{A_{k,n}^u}$ is the first-order partial derivative of $A_{k,n}^u =  \sqrt{{\Pr}^{u,\textrm{LoS}}_{k,n}}  \boldsymbol{\varphi}_k^{u,\textrm{LoS}} [n]+ \sqrt{{\Pr}^{u,\textrm{NLoS}}_{k,n} \kappa} \boldsymbol{\varphi}_k^{u,\textrm{NLoS}} [n] $, w.r.t $\phi_{k,n}^u$.

\subsection{UAV Energy Consumption Model}
As UAVs are energy-constrained devices, in this paper, we take into account the energy consumption of UAVs. The energy consumption at each UAV includes two parts: communication and sensing energy and flying energy. Thus, the communication and sensing energy at UAV $u$ at time slot $n$ can be formulated as:
\begin{equation}
        E^{u,\textrm{cs}} [n] = \tau \left( \sum_{v=1}^{V_u}\left\lVert\boldsymbol{g}_v^u[n]\right\lVert^2  + \sum_{k=1}^{K_u}\left\lVert\boldsymbol{i}_k^u[n]\right\lVert^2  \right), \forall u,\forall n. 
 \end{equation}
Then, the flying energy of UAV $u$ at time slot $n$ can be expressed as:
\begin{equation}
\begin{split}
     E^{u,\textrm{fl}}[n] &= \tau\Biggr[C_0\bigg(1 + \frac{3(a_u^h[n])^2}{U_{\mathrm{tip}}^2}\bigg) + \frac{1}{2}\psi_0 \Tilde{r}\rho G(a_u^h[n])^3 +  \\ 
     & C_1\bigg(\sqrt{1 + \frac{(a_u^h[n])^4}{4a_o^4}} - \frac{(a_u^h[n])^2}{2a_0^2}\bigg)^{\frac{1}{2}} + C_2a_u^c[n]\Biggr],\forall u,n,
\end{split}
\end{equation} 
where the constants $C_0$ and $C_1$ represent the induced power and blade profile power in the hovering state, respectively. The rotor blade's tip speed is $U_{\mathrm{tip}}$, and the constant for descending and ascending power is $C_2$. The rotor solidity and fuselage drag ratio are denoted by $\Tilde{r}$ and $\psi_0$, respectively. The air density and rotor disc area are denoted by $\rho$ and $G$, respectively. The hovering velocity is denoted by $a_0$, while the vertical velocity of UAV $u$ is represented by $a_u^c[t]=\frac{|H_u[n] - H_u[n-1]|}{\tau}$.
\section{Problem Formulation and Solution Approach}
Our objective is to jointly optimize the beamforming vectors of communication and sensing and the trajectories of UAVs to maximize the communication rate while guaranteeing sensing performance. We formally pose the proposed problem as:
	\begin{maxi!}[2]                 
		{\boldsymbol{g,i, o}}                               
		{\sum_{n=1}^{N}\sum_{u=1}^{U}\sum_{v=1}^{V_u} R_v^u[n]}{\label{opt:P}}{\textbf{P:}} 
		\addConstraint{\text{CRB} (\phi_{k,n}^u) \leq \Gamma_{k,n}^{u, \textrm{sen}}, \forall u, \forall k, \forall n,}    \label{8b}
		\addConstraint{\sum_{v=1}^{V_u} \left\lVert\boldsymbol{g}_v^u[n]\right\lVert^2  + \sum_{k=1}^{K_u}\left\lVert\boldsymbol{i}_k^u[n]\right\lVert^2 \leq P_u^{\textrm{max}}, \forall u, \forall n,} \label{8c}
		\addConstraint{\sum_{n=1}^{N}  E^{u,\textrm{cs}} [n] + \sum_{n=1}^{N}  E^{u,\textrm{fl}}[n] \leq E^{u, \textrm{th}}, \forall u,} \label{8d}
		\addConstraint{\left\lVert o_u[n] - o_u[n-1] \right\lVert \leq a_{\textrm{max}}\tau, \forall u, \forall n,} \label{8e}
        \addConstraint{o_u[0] = o_0, o_u[N] = o_f, \forall u,}  \label{8f}
        \addConstraint{H_{\textrm{min}}\leq H_u[n] \leq H_{\textrm{max}}, \forall u, \forall n, } \label{8g}
        \addConstraint{\left\lVert o_u[n] - o_{u'}[n]\right\lVert \geq d_{\textrm{min}}, \forall u,u' \in \mathcal{U}, u \neq u'}, \label{8h}
  \vspace{-0.1in}
	\end{maxi!}
where constraint (\ref{8b}) guarantees the sensing performance, and constraints (\ref{8c}) and (\ref{8d}) present the power budget and energy limit of each UAV. Constraints (\ref{8e}) and (\ref{8f}) ensure the velocity constraint and the initial and final locations of the UAV, respectively. Finally, constraints (\ref{8g}) and (\ref{8h}) ensure the maximum altitude of each UAV to maintain reliable communication and sensing links and collision avoidance between UAVs, respectively. Due to the non-convex objective function and constraints, our proposed problem in \textbf{P} is the non-convex problem. Thus, using the BCD method, we decompose the problem into two subproblems: the beamforming design problem for communication and sensing and the UAV trajectory optimization problem. Then, we propose a fractional programming approach and the DDPG method to solve the decomposed subproblems alternatively. 
 \vspace{-0.2in}
\subsection{Beamforming Design Problem for Communication and Sensing}
We introduce $\boldsymbol{G}_v^u[n] = \boldsymbol{g}_v^u [n] (\boldsymbol{g}_v^u [n])^\dagger$, and $\boldsymbol{I}_k^u[n] = \boldsymbol{i}_k^u [n] (\boldsymbol{i}_k^u [n])^\dagger$, where $\boldsymbol{G}_v^u[n], \boldsymbol{I}_k^u[n] \geq 0, \text{rank}(\boldsymbol{G}_v^u[n])= \text{rank}(\boldsymbol{I}_k^u[n]) =1$, $\mathcal{H}_v^u[n] = \boldsymbol{h}_v^u[n] (\boldsymbol{h}_v^u[n])^\dagger$, $\mathcal{R}(\boldsymbol{G}_v^u[n], \boldsymbol{I}_k^u[n]) = \tau \log_2\left(1+ \frac{\Tr(\mathcal{H}_v^u[n]\boldsymbol{G}_v^u[n])}{\Theta_v^u(\boldsymbol{G}_v^u[n], \boldsymbol{I}_k^u[n])} \right)$, where $\Theta_v^u(\boldsymbol{G}_v^u[n], \boldsymbol{I}_k^u[n]) = \sum_{v'=1, v' \neq v }^{V_u} \Tr(\mathcal{H}_{v}^u[n] \boldsymbol{G}_{v'}^{u}[n]) + \sum_{k=1}^{K_u} \Tr(\mathcal{H}_{v}^u[n]\boldsymbol{I}_{k}^u[n]) +  \sum_{u'=1, u' \neq u}^{U}\sum_{v=1}^{V_{u'}} \Tr(\mathcal{H}_{v}^{u'}[n]\boldsymbol{G}_{v'}^{u'}[n]) + \sum_{u'=1, u' \neq u}^{U} \sum_{k=1}^{K_{u'}} \Tr(\mathcal{H}_{v}^{u'}[n] \boldsymbol{I}_{k}^{u'}[n]) + \sigma^2$. Then, at the given UAV trajectory, we formulate the beamforming design problem as:
\begin{maxi!}[2]                 
		{\boldsymbol{G, I}}                               
		{\sum_{n=1}^{N}\sum_{u=1}^{U}\sum_{v=1}^{V_u} \mathcal{R}_v^u[n] (\boldsymbol{G,I}) }{\label{opt:P1}}{\textbf{P1:}} 
		\addConstraint{\Tr((\overline{A_{k,n}^u})^\dagger\overline{A_{k,n}^u}\boldsymbol{I}_k^u[n] ) \geq \frac{\sigma^2}{2\Gamma_{k,n}^{u, \textrm{sen}}|\beta_{k,n}^u|^2} ,} \nonumber
        \addConstraint{ \ \ \ \ \ \ \ \ \ \ \ \ \ \ \ \ \ \ \ \ \ \ \ \ \ \ \   \forall u, \forall k, \forall n,}
	\addConstraint{\sum_{v=1}^{V_u} \Tr(\boldsymbol{G}_v^u[n])  + \sum_{k=1}^{K_u}\Tr(\boldsymbol{I}_k^u[n])\leq P_u^{\textrm{max}},} \nonumber
 \addConstraint{ \ \ \ \ \ \ \ \ \ \ \ \ \ \ \ \ \ \ \ \ \ \ \ \ \ \ \  \ \ \ \ \ \ \  \forall u, \forall n,}
 \addConstraint{\sum_{n=1}^{N} \tau\left(\sum_{v=1}^{V_u} \Tr(\boldsymbol{G}_v^u[n])  + \sum_{k=1}^{K_u}\Tr(\boldsymbol{I}_k^u[n])\right) } \nonumber
 \addConstraint{ \ \ \ \  +\sum_{n=1}^{N}  E^{u,\textrm{fl}}[n] \leq E^{u, \textrm{th}}, \forall u,  }
 \addConstraint{\boldsymbol{G}_v^u[n], \boldsymbol{I}_k^u[n] \succeq 0, \forall u, \forall v, \forall k, \forall n,}
 \addConstraint{\text{rank}(\boldsymbol{G}_v^u[n])= \text{rank}(\boldsymbol{I}_k^u[n]) =1,\forall u,v,k,n,}
	\end{maxi!}
where \textbf{P1} is non-convex due to the strict coupling of optimization variables in both objective function and constraints (9c) and (9d). Thus, we decompose  \textbf{P1} into a beamforming design problem for communication and a beamforming design problem for sensing.
\subsubsection{Beamforming Design Problem for Communication}
At the given $\boldsymbol{o}$, and $\boldsymbol{I}$, we can formulate the beamforming design problem for communication as: 
\begin{maxi!}[2]                 
		{\boldsymbol{G}}                               
		{\sum_{n=1}^{N}\sum_{u=1}^{U}\sum_{v=1}^{V_u} \mathcal{R}_v^u[n] (\boldsymbol{G}) }{\label{opt:P2}}{\textbf{P2:}} 
		\addConstraint{\text{(9c)}, \text{(9d)},}
        \addConstraint{\boldsymbol{G}_v^u[n] \succeq 0, \forall u, \forall v, \forall n,}
          \addConstraint{\text{rank}(\boldsymbol{G}_v^u[n]) =1, \forall u, \forall v, \forall n.}
	\end{maxi!}
Then, by using Lagrangian dual transformation method \cite{shen2018fractional}, we can reformulate \textbf{P2} as:   
\begin{maxi!}[2]                 
		{\boldsymbol{G, \chi}}                               
		{\sum_{n=1}^{N}\sum_{u=1}^{U}\sum_{v=1}^{V_u} \tau(\log_2(1+\chi_v^u[n]) - \chi_v^u[n]) + \frac{O_v^u[n]}{P_v^u[n]}}{\label{opt:P3}}{\textbf{P3:}} 
		\addConstraint{\text{(9c)}, \text{(9d)},\text{(10c)}, \text{(10d)}, }
	\end{maxi!}
where $O_v^u[n] = \tau(1+\chi_v^u[n])\Tr(\mathcal{H}_v^u[n]\boldsymbol{G}_v^u[n]) $, and $P_v^u[n] = \Tr(\mathcal{H}_v^u[n]\boldsymbol{G}_v^u[n])+ \Theta_v^u(\boldsymbol{G}_v^u[n], \boldsymbol{I}_k^u[n])$. Furthermore, $\boldsymbol{\chi}$ denotes the set of auxiliary variables, where the optimal value of an auxiliary variable, $\chi_v^{u*}[n] = \frac{\Tr(\mathcal{H}_v^u[n]\boldsymbol{G}_v^u[n])}{\Theta_v^u(\boldsymbol{G}_v^u[n], \boldsymbol{I}_k^u[n])}, \forall v,u,n$. Furthermore, by using the quadratic transform method \cite{shen2018fractional}, we can reformulate \textbf{P3} into a more tractable form as follows:
\begin{maxi!}[2]                 
		{\boldsymbol{G, \chi, \psi }}                               
		{\sum_{n=1}^{N}\sum_{u=1}^{U}\sum_{v=1}^{V_u} \tau(\log_2(1+\chi_v^u[n]) - \chi_v^u[n]) + \Psi_v^u[n] }{\label{opt:P4}}{\textbf{P4:}} 
		\addConstraint{\text{(9c)}, \text{(9d)},\text{(10c)}, \text{(10d)}, }
	\end{maxi!}
where $\Psi_v^u[n]= 2 \psi_v^u[n] \sqrt{O_v^u[n]}- (\psi_v^u[n])^2P_v^u[n]$, and $\psi_v^{u*}[n] = \frac{\sqrt{O_v^u[n]}}{P_v^u[n]}$. However, because of the non-convex rank constraints in (10d), \textbf{P4} is still non-convex. We relaxed the rank constraints, which turned \textbf{P4} over $\boldsymbol{G}$ into a conventional convex optimization problem that can be solved optimally, as shown in Algorithm 1. For the rank constraints, it is possible to determine that there is always an optimal rank-one solution $\boldsymbol{G^*}$ by examining the Karush-Kuhn-Tucker (KKT) optimality conditions of the relaxed version of problem \textbf{P4} and using an analysis akin to that found in the Appendix of \cite{mu2021simultaneously}. 
\begin{algorithm}[t]
\caption{\strut Beamforming Design for Communication}
\label{alg:comm_BF}
\textbf{Initialization:} Set $\hat{t}=0$, and find initial feasible solutions $(\boldsymbol{G}^{0}, \boldsymbol{\chi}^{0}, \boldsymbol{\Psi}^{0})$; \\
\textbf{repeat} \\
\hspace{0.5cm} At the given $\boldsymbol{G}^{\hat{t}}$ and $\boldsymbol{\Psi}^{\hat{t}}$, update the value of auxiliary variable, $\boldsymbol{\chi}^{\hat{t}+1}$;   \\
\hspace{0.5cm} At the given $\boldsymbol{G}^{\hat{t}}$ and $\boldsymbol{\chi}^{\hat{t}+1}$, update the value of auxiliary variable, $\boldsymbol{\Psi}^{\hat{t}+1}$;  \\
\hspace{0.5cm} At the given $\boldsymbol{\chi}^{\hat{t}+1}$ and $\boldsymbol{\Psi}^{\hat{t}+1}$, update the value of auxiliary variable, $\boldsymbol{G}^{\hat{t}+1}$;    \\
\hspace{0.5cm} Update $\hat{t}= \hat{t} + 1$; \\
\textbf{until} The objective function converges;  \\
\textbf{Then,} set $\boldsymbol{\chi}^{\hat{t}+1}$, $\boldsymbol{\Psi}^{\hat{t}+1}$, and  $\boldsymbol{G}^{\hat{t}+1}$ as the desired solution.
\end{algorithm}
\subsubsection{Beamforming Design Problem for Sensing}
At the given $\boldsymbol{o}$, and $\boldsymbol{G}$, we can formulate the beamforming design problem for sensing as:
\begin{maxi!}[2]                 
		{\boldsymbol{I}}                               
		{\sum_{n=1}^{N}\sum_{u=1}^{U}\sum_{v=1}^{V_u} \mathcal{R}_v^u[n] (\boldsymbol{I}) }{\label{opt:P5}}{\textbf{P5:}} 
		\addConstraint{\text{(9b)}, \text{(9c), \text{(9d)}}},
        \addConstraint{\boldsymbol{I}_k^u[n] \succeq 0, \forall u, \forall k, \forall n,}
          \addConstraint{\text{rank}(\boldsymbol{I}_k^u[n]) =1, \forall u, \forall k, \forall n.}
	\end{maxi!}
However, \textbf{P5} is non-convex due to non-convex objective function and rank one constraints. Thus, to tackle \textbf{P5}, we first introduce $\iota_v^u[n] =  \frac{\Tr(\mathcal{H}_v^u[n]\boldsymbol{G}_v^u[n])}{\Theta_v^u(\boldsymbol{G}_v^u[n], \boldsymbol{I}_k^u[n])}, \forall v,u,n$, and then we can reformulate $\textbf{P5}$ as:
\begin{maxi!}[2]                 
		{\boldsymbol{I,\iota }}                               
		{\sum_{n=1}^{N}\sum_{u=1}^{U}\sum_{v=1}^{V_u} \tau \log_2(1+\iota_v^u[n])}{\label{opt:P6}}{\textbf{P6:}} 
		\addConstraint{\text{(9b)}, \text{(9c)}, \text{(9d)}, \text{(13c)}, \text{(13d)},}
        \addConstraint{\Tr(\mathcal{H}_v^u[n]\boldsymbol{G}_v^u[n]) \geq \iota_v^u[n] \Theta_v^u(\boldsymbol{G}_v^u[n], \boldsymbol{I}_k^u[n]).}
	\end{maxi!}
Then, we introduce the upper-bound approximation function of the non-convex constraint (14c) as follows:
\begin{equation}
    \Tr(\mathcal{H}_v^u[n]\boldsymbol{G}_v^u[n]) \geq \frac{\big(\Theta_v^u(\boldsymbol{G}_v^u[n], \boldsymbol{I}_k^u[n])\big)^2}{2\Omega_v^u(\boldsymbol{G}_v^u[n], \boldsymbol{I}_k^u[n])} + \frac{\big(\iota_v^u[n]\big)^2\Omega_v^u(\boldsymbol{G}_v^u[n], \boldsymbol{I}_k^u[n]) }{2},
\end{equation}
where the condition always holds if $\Omega_v^u(\boldsymbol{G}_v^u[n], \boldsymbol{I}_k^u[n]) = \Theta_v^u(\boldsymbol{G}_v^u[n], \boldsymbol{I}_k^u[n])/\iota_v^u[n]$. Then, we can reformulate \textbf{P6} as:
\begin{maxi!}[2]                 
		{\boldsymbol{I,\iota }}                               
		{\sum_{n=1}^{N}\sum_{u=1}^{U}\sum_{v=1}^{V_u} \tau \log_2(1+\iota_v^u[n])}{\label{opt:P7}}{\textbf{P7:}} 
		\addConstraint{\text{(9b)}, \text{(9c)}, \text{(9d)}, \text{(13c)}, \text{(13d)}, \text{(15)}.}
	\end{maxi!}
Finally, we relaxed the rank constraints, turning \textbf{P7} over $\boldsymbol{I,\iota }$ into a conventional convex optimization problem CVX can solve optimally, as shown in Algorithm 2. 
\begin{algorithm}[t]
\caption{\strut Beamforming Design for Sensing}
\label{alg:sens_BF}
\textbf{Initialization:} Set $\hat{t}=0$, and find initial feasible solutions $(\boldsymbol{I}^{0}, \boldsymbol{\iota}^{0})$;  \\
\textbf{repeat}   \\
\hspace{0.5cm} At the given $\boldsymbol{I}^{\hat{t}}$ and $\boldsymbol{\iota}^{\hat{t}}$, update $\boldsymbol{\Omega}^{\hat{t}+1}$;  \\
\hspace{0.5cm} At the given $\boldsymbol{\Omega}^{\hat{t}+1}$, update $\boldsymbol{I}^{\hat{t}+1}$ and $\boldsymbol{\iota}^{\hat{t}+1}$;    \\
\hspace{0.5cm} Update $\hat{t} = \hat{t} + 1$; \\
\textbf{until} The objective function converges; \\
\textbf{Then,} set $\boldsymbol{I}^{\hat{t}+1}$ and $\boldsymbol{\iota}^{\hat{t}+1}$ as the desired solution.
\end{algorithm}

\subsubsection{UAV Trajectory Optimization}
At the given $\boldsymbol{G}$, and $\boldsymbol{I}$, we can formulate the UAV trajectory optimization problem as:
\begin{maxi!}[2]                 
		{\boldsymbol{o}}                               
		{\sum_{n=1}^{N}\sum_{u=1}^{U}\sum_{v=1}^{V_u} \mathcal{R}_v^u[n] (\boldsymbol{o})}{\label{opt:P8}}{\textbf{P8:}} 
		\addConstraint{\text{(8b)}, \text{(8d)}, \text{(8e),} \text{(8f)}, \text{(8g)}, \text{(8h).}}  \label{17b}
\end{maxi!}
As we can see from \textbf{P8}, the objective function and constraints (\ref{8b}), (\ref{8d}), (\ref{8e}), and (\ref{8h}) are non-convex with respect to $\boldsymbol{o}$. Moreover, the network condition is highly dynamic due to the mobility of UAVs. To address those challenges, we exploit the DDPG algorithm to determine the trajectories of UAVs. Firstly, to cast \textbf{P8} into the Markov decision process, we define state, action, and reward as follows:
\begin{itemize}
    \item \textit{State:} The state space $s(n)$ of the system consists of the communication and sensing beamforming information, the locations of communication users and sensing targets, and the positions of UAVs in the current time slot.
    \item \textit{Action:} The action space is defined as the trajectory of UAVs (i.e., the horizontal flying velocity $a_u^h[n]$, directional angle $\theta_u[n]$, and the height $H_u[n]$). Therefore, the action space is defined as $a(n) = \{a_u^h[n], \theta_u[n], H_u[n], \forall u \in \mathcal{U}\}$.
    \item \textit{Reward:} We define the reward function for {\textbf{P8}} as
\end{itemize}
\begin{equation}
    r(n) = \sum_{u=1}^{U}\sum_{v=1}^{V_u} \mathcal{R}_v^u[n] (\boldsymbol{o}) - \xi,  \label{reward_eq}
\end{equation}
where $\xi$ is the penalty term for the deviation of any one of the constraints given in (\ref{17b}). Algorithm \ref{alg:ddpg_table} presents a detailed description of the proposed trajectory optimization.
\begin{algorithm}[t!]  
\SetAlgoLined
\caption{DDPG-Based Multi-UAV Trajectory Optimization}
\label{alg:ddpg_table}
\textnormal{
Initialize the parameters $\nu^\mu$ and $\nu^{\mu'}$ of the policy network $\mu(.)$ and target policy network $\mu'(.)$, and set $\nu^{\mu'} = \nu^\mu$; \\
Initialize the parameters $\nu^Q$ and $\nu^{Q'}$ of the Q-network $Q(.)$ and target Q-network $Q'(.)$, respectively, and set $\nu^{Q'} = \nu^Q$;  \\
\For{episode $= 1$ to $E$}{
    Initialize the random process $\Omega$ for the exploration of the action; \\
    Observe state $s(1)$; \\
    \For{$n = 1$ to $N$}{
    \While{(\ref{8b}), (\ref{8d})-(\ref{8h}) are not satisfied and until the maximum step has reached}{
         According to the current policy and exploration noise, select the action $a(n) = \mu\big(s(n)|\nu^\mu\big) + \Omega_n$; \\
        \If{The constraints are deviated}{
            Cancel the action and apply the penalty; 
        }
        Store the transition $(s(n), a(n), r\big(s(n), a(n)\big), s(n+1))$ in experience replay buffer $B$; 
  }
    Random minibatch of $H$ transitions $(s_i, a_i, r_i, s_{i+1})$ from $B$ are sampled; \\
    Compute the target Q-value $y(i)=$
    \vspace{-0.2cm}
               \[r(i) + \gamma Q'\bigg(s(i+1), \mu'\big(s(i+1)|\nu^{\mu'}\big)|\nu^{Q'}\bigg); \] \\
    Update the weights $\theta^Q$ of $Q(.)$ by minimizing the loss function
            \[
            L(\theta^Q) = \frac{1}{H}\sum_{i=1}^{H}\bigg(y(i)-Q\big(s(i), a(i)|\nu^Q\big)\bigg);
            \]    \\
    Update the target Q-network parameters: $\nu^{Q'} = \delta\nu^Q + (1-\delta)\nu^{Q'}$; \\
    Update the weights $\nu^\mu$ of the policy network $\mu(.)$ by applying the gradient method:
        $\nabla_{\nu^\mu}J(\nu^{\mu}) \approx $
        \[
         \frac{1}{H}\sum_{i}^{H}\nabla_{a} Q(s, a|\nu^Q)|_{s=s(i), a=\mu(s(i))}  
         \nabla_{\nu^\mu}\mu(s|\nu^\mu)|_{s(i)};\] \\
    Update the target policy network: $\nu^{\mu'}= \delta\nu^\mu + (1 - \delta)\nu^{\mu'}$; \\
    Obtain the optimal flying velocity, directional angle, and height of the UAVs; \\
    Evaluate the position of UAVs by \\
    $x_u[n+1] = x_u[n] + \tau a_u^h[n]\cos{(\theta_u[n])}$, \\
    $y_u[n+1] = y_u[n] + \tau a_u^h[n]\sin{(\theta_u[n])}$;
    }
}
}
\end{algorithm}
\begin{table}[t] 
	\caption{Simulation parameters.}
	\label{fig:Table}
	\setlength{\tabcolsep}{3pt} 
	\renewcommand{\arraystretch}{0.9} 
	\scriptsize 
	\begin{center}
		\begin{tabular}{|p{1.4cm}|p{2cm}|p{1.4cm}|p{2cm}|} 
			\hline 
			\textbf{Parameter} & \textbf{Value} & \textbf{Parameter} & \textbf{Value}  \\ \hline  \hline
			$U, M$ & $3, 3$ & $g_0$ & $-70$ dB  \\ \hline 
			$V_u$ & $[3,5]$ & $C, D$ & $11.95, 0.136$  \\ \hline  
			$K_u$ & $[2,4]$ & $\theta_u[n]$ & $[\frac{-5\pi}{12}, \frac{5\pi}{12}]$  \\ \hline 
			$a_u^h[n]$ & $[10,20]$ m/s & $B$, batch size  & $1600, 32$ \\ \hline
			$H_u[n]$  & $[150, 200]$ m & $U_{\textrm{tip}}, \psi_0$ & $120, 0.6$    \\ \hline
			$C_0, C_1$  & $798.6, 88.6$ & $\Tilde{r}, \rho$ & $0.005, 1.226$  \\ \hline
			$C_2, \sigma_k$ & $11.5, -17$ dBsqm & $G, a_0$ & $0.503, 4.3$     \\ \hline
		\end{tabular}
	\end{center}
\end{table}
\begin{figure*}[t] 
    \centering
    \subfloat[Trajectories of UAVs.]{%
        \includegraphics[width=0.43\textwidth]{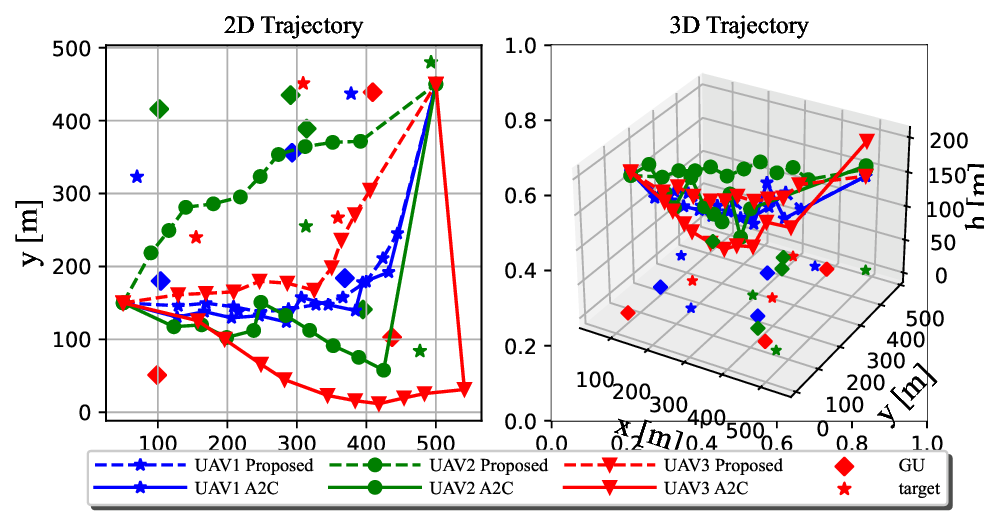}%
        \label{fig:traj}%
    }
    \subfloat[Sum rate vs CRB.]{%
        \includegraphics[width=0.27\textwidth]{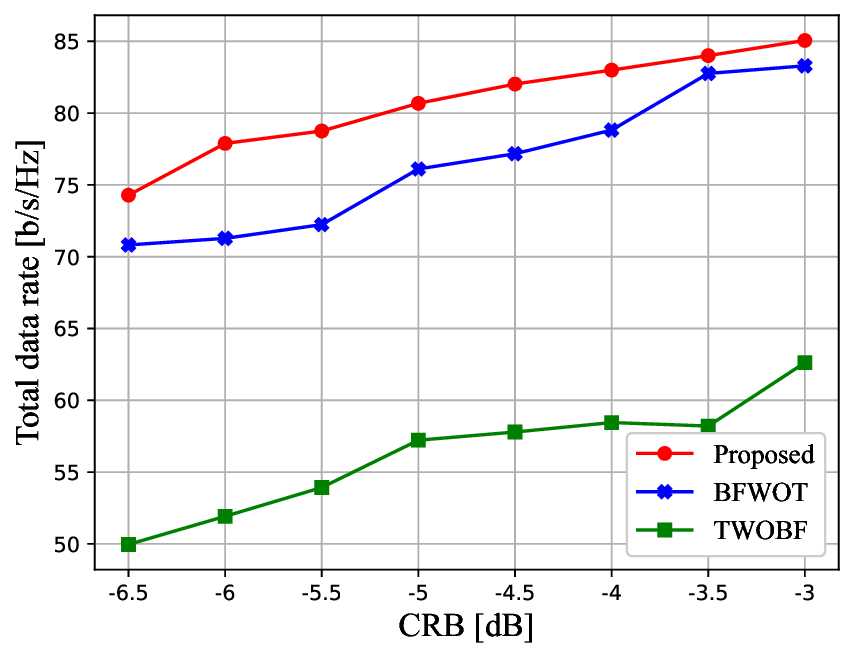}%
        \label{fig:rate_crb}%
    }
    \subfloat[Convergence.]{%
        \includegraphics[width=0.27\textwidth]{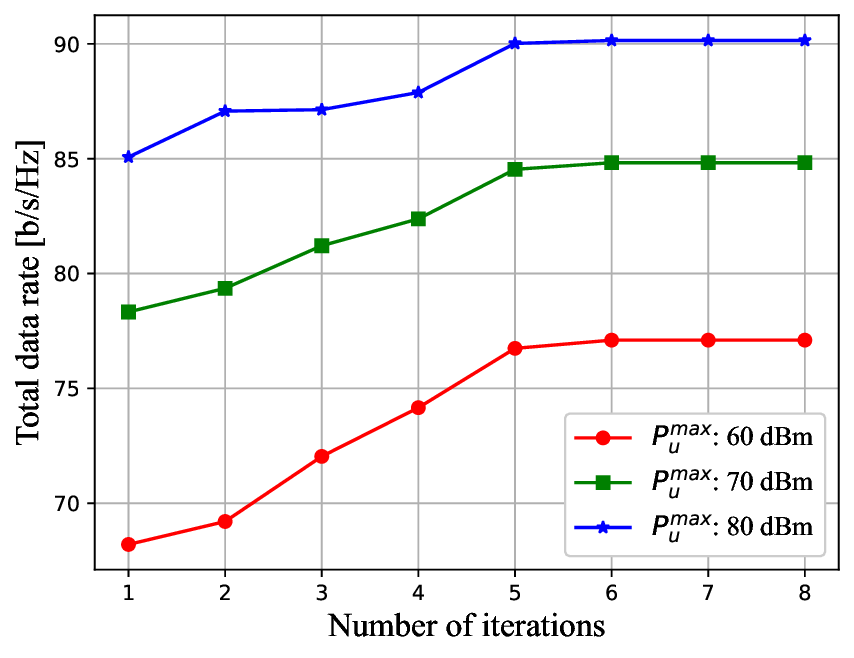}%
        \label{fig:convergence}%
    }
    \caption{(a) Comparison of UAV trajectories. (b) Rate vs CRB. (c) Convergence of the proposed algorithm in the ISAC system.}
    \label{fig:combined_figures}
\end{figure*}
\section{Simulation Results and Analysis}
To evaluate the proposed algorithm, we consider a square area of $500$ m $\times$ $500$ m, where all UAVs share identical starting and destination points. Unless specified otherwise, the parameters used in the simulations are listed in Table \ref{fig:Table}.

In Fig. \ref{fig:traj}, we illustrate the UAV trajectories in both 2D and 3D perspectives, comparing our proposed DDPG-based approach with the A2C algorithm. As depicted, the UAVs in our approach maintain proximity to their respective users while traversing from the starting to the destination point. This behaviour highlights its superior path-planning capability, 
attributed to the higher sample efficiency and improved policy update mechanism of DDPG. In contrast, the A2C algorithm, limited by lower sample efficiency and reliance on current data for policy updates, results in UAV paths that are relatively distant from the users. For instance, the trajectories of UAV2 and UAV3 under A2C significantly deviate from user locations. The variation in UAV altitudes for both approaches is also presented, demonstrating the adaptability of our method.

Fig. \ref{fig:rate_crb} illustrates the impact of the CRB on the total data rate of the system. We compare our approach against two baselines to validate the performance gains, i.e., TWOBF (trajectory without beamforming) and BFWOT (beamforming without trajectory). The results indicate that TWOBF achieves the lowest total data rate due to suboptimal resource allocation, which leads to increased interference from sensing signals and inefficient power distribution. This interference degrades the SINR of communication users. As the CRB value increases, the sensing signal strength diminishes, introducing less interference to communication signals and thereby enhancing the total data rate. The superior performance of our proposed algorithm across varying CRB values demonstrates its capability to effectively balance sensing and communication resources.

Fig. \ref{fig:convergence} presents the convergence behaviour of our proposed algorithm under different power thresholds. As expected, higher power levels result in greater total data rates. Notably, our algorithm converges to a stationary point within six iterations, affirming its practical feasibility and computational efficiency. This rapid convergence is indicative of the algorithm's potential for real-time applications in ISAC systems. Our simulation results demonstrate the efficacy of the proposed DDPG-based approach in optimizing UAV trajectories, enhancing data rates, and ensuring rapid convergence, thereby making a valuable contribution to the field of ISAC systems.

\section{Conclusion} 
In this paper, we have presented a joint beamforming and trajectory optimization approach for a multi-UAV-assisted ISAC system to improve both communication throughput and sensing accuracy. Our proposed BCD-based iterative algorithm, which combines fractional programming and a DDPG-based algorithm, was shown to be highly effective through simulation results. It significantly improved the sum rate of users while maintaining reliable target sensing. Our proposed approach has demonstrated superior performance by efficiently coordinating beamforming and UAV trajectories compared to conventional separated frameworks. Moreover, the results emphasized the balancing of sensing and communication objectives through an effective multi-UAV coordination. These findings highlight the potential of the proposed method for advancing ISAC systems in real-world applications.  
\bibliographystyle{IEEEtran}
\bibliography{Globecom}
\end{document}